\documentclass{ws-ijmpcs}

\begin{document}

\markboth{Analisa Mariazzi for the Pierre Auger Collaboration}
{Highest energy particle physics with the Pierre Auger Observatory}

%
\catchline{}{}{}{}{}
%

\title{HIGHEST ENERGY PARTICLE PHYSICS WITH THE PIERRE AUGER OBSERVATORY
}

\author{ANALISA MARIAZZI
}
\address{IFLP, Universidad Nacional de La Plata and CONICET, La Plata(1900), Argentina
mariazzi@fisica.unlp.edu.ar}

\author{for the Pierre Auger Collaboration}

\address{Observatorio Pierre Auger, Av. San Martin Norte 304, 5613 Malarg\"ue, Argentina\footnote{
Full author list: http://www.auger.org/archive/authors\_2013\_05.html}\\
auger\_spokespersons@fnal.gov}

\maketitle

\begin{history}
\received{1 11 2013}
\revised{5 11 2013}
\end{history}

\begin{abstract}
Astroparticles offer a new path for research in the field of particle physics, allowing investigations 
at energies above those accesible with accelerators.
Ultra-high energy cosmic rays can be studied via the observation of the showers they 
generate in the atmosphere. 
The Pierre Auger Observatory is a hybrid detector for ultra-high energy cosmic rays, combining 
two complementary measurement techniques used by previous experiments, to get the best 
possible measurements of these air showers. Shower observations enable one to 
not only estimate the energy, direction and most probable mass of the primary cosmic 
particles but also to obtain some information about the properties of their hadronic interactions. 
Results that are most relevant in the context of determining hadronic interaction characteristics at ultra-high energies 
will be presented.
\end{abstract}
\ccode{PACS numbers: 
96.50.S-  
96.50.sd  
96.50.sb  
}
\section{Introduction}
Cosmic rays from astrophysical sources provide a natural beam of ultra high energy
particles that can be used to probe particle interactions at the highest energies.

The interpretation of cosmic ray measurements requires modeling of hadronic interactions 
in an energy range beyond that which can be studied in accelerator experiments. 
The knowledge of the relevant properties of hadronic interactions in this energy range 
is therefore of central importance for the interpretation of the cosmic ray data. 
Nevertheless, it is in principle possible to obtain information about hadronic 
interactions from the cosmic ray observations, but dealing with the fact that 
this natural cosmic ray beam has an unknown energy spectrum and an unknown mass 
composition. 
The mass composition of the primary particles must be estimated from the same data set.
Solving the ambiguity between composition and hadronic interaction modeling is a
key problem for ultra-high energy cosmic ray observations. 

The Pierre Auger Observatory, the world\'s largest cosmic ray observatory, is
located near Malarg\"{u}e, in the Province of Mendoza, Argentina. It was designed 
to investigate the origin and the nature of ultra-high energy cosmic rays by
taking advantage of two available techniques to detect extensive air showers initiated by 
ultra-high energy cosmic rays: a surface detector (SD) array and a fluorescence detector (FD). 
The SD consists of an array of about 1600 water-Cherenkov surface detectors deployed over a
triangular grid of 1.5 km spacing and covering an area of 3000 km$^2$. 
The SD is overlooked by 27 fluorescence telescopes, grouped in four sites, making up the 
fluorescence detector.
The FD observes the longitudinal development of the shower in the atmosphere 
by detecting the fluorescence light emitted by excited 
nitrogen molecules and Cherenkov light induced by shower particles in air. 
The FD provides a calorimetric measurement of the primary particle energy. 

These two detection methods are complementary, so that combining them in hybrid mode
will help resolve mass composition and hadronic interaction information.  
\section{Proton-air cross section from air showers}
\begin{figure}[pb]
\centerline{
\includegraphics[width=6.cm]{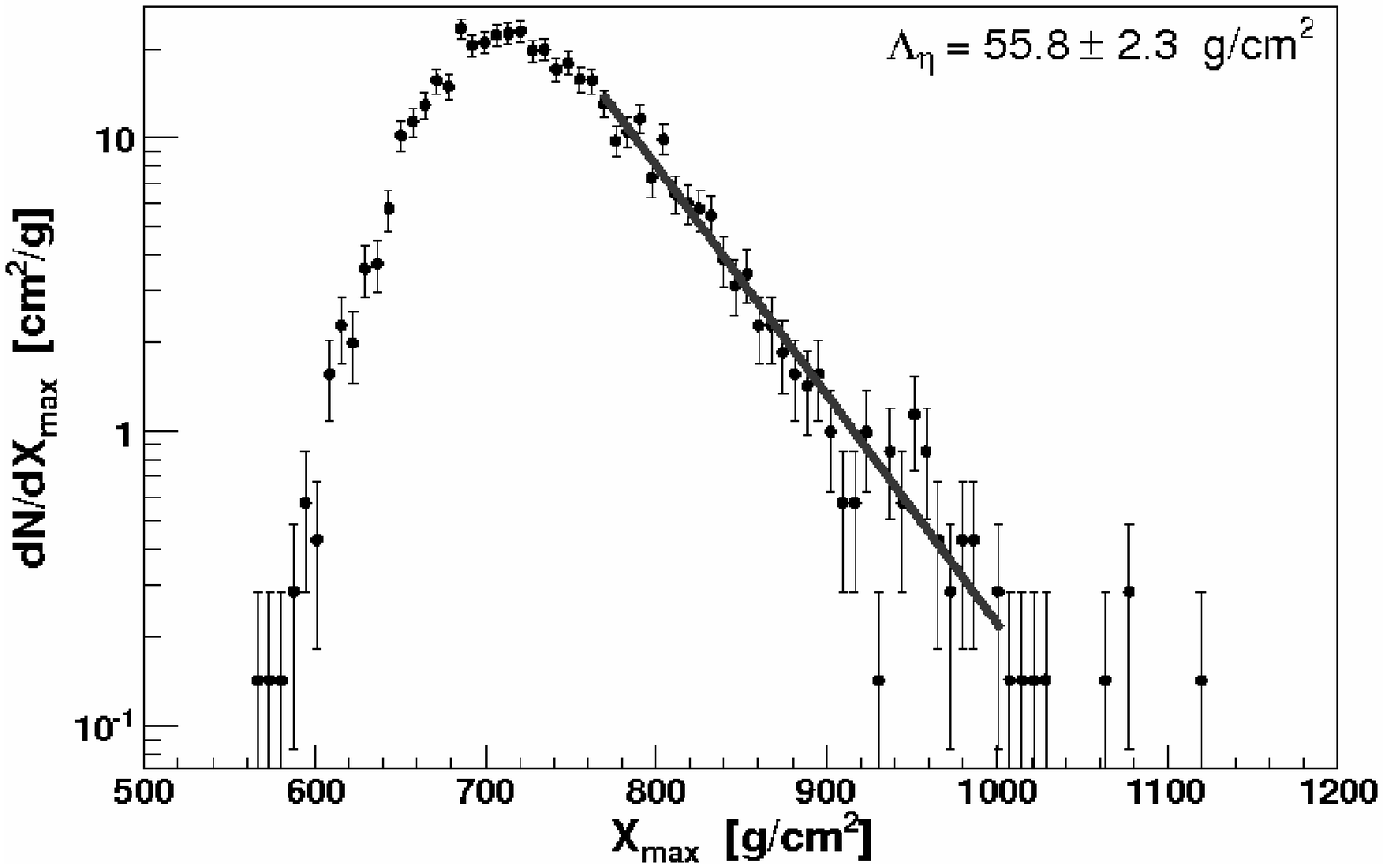}
\includegraphics[width=6.cm]{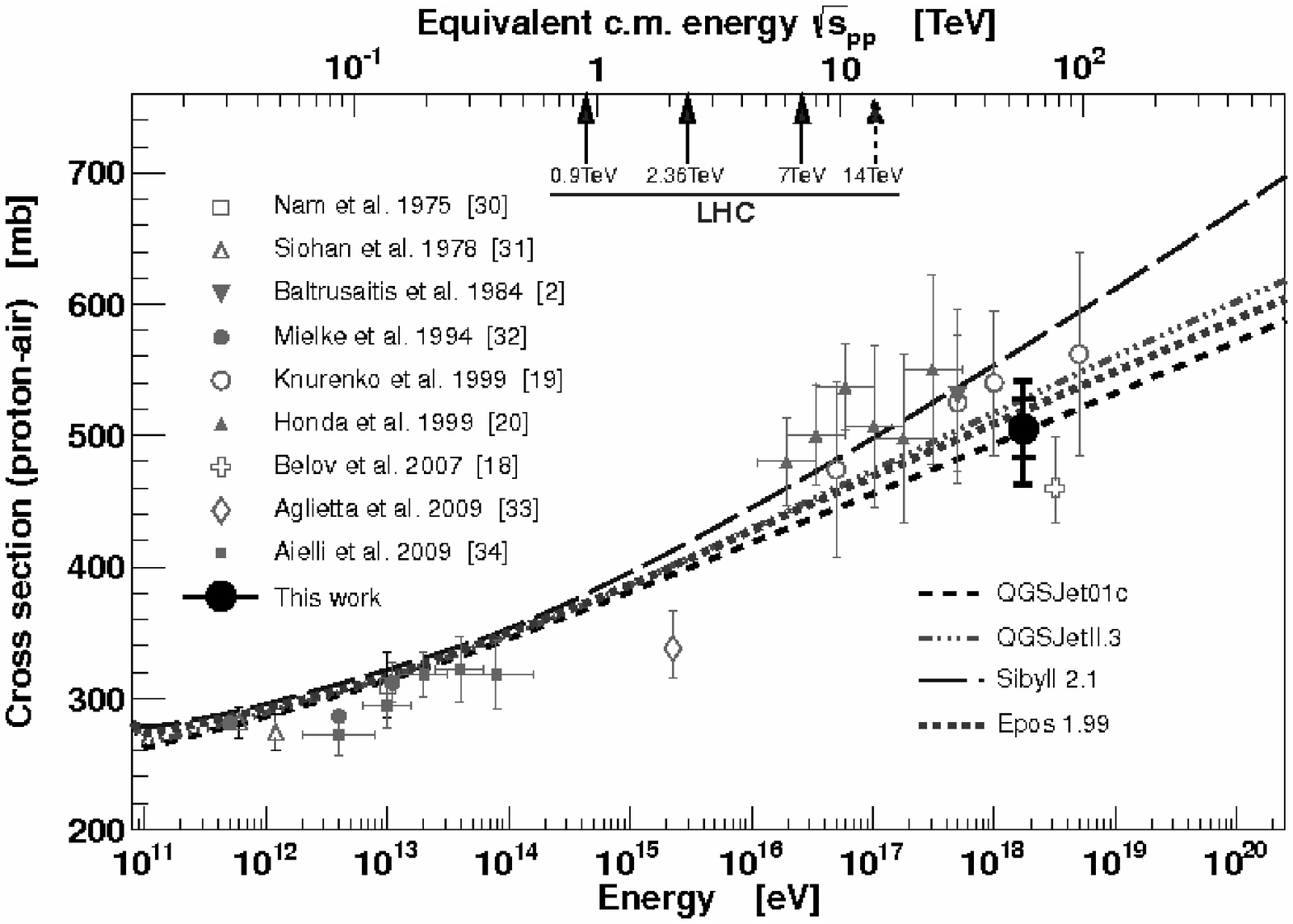} 
           }
\vspace*{8pt}
\caption{(left)-Fit to the tail of the $X_{max}$ distribution.
(right)- Resulting inelastic proton-air cross-section compared to other measurements and model predictions. \label{crossection}}
\end{figure}
The tail of the $X_{max}$ (depth at which the shower reaches its maximum size) 
distributions, is shown to be highly sensitive to cross section.
Considering only the most deeply penetrating air showers enhances the proportion of protons 
in a sample as the average depth of shower maximum is higher in the atmosphere 
for heavier primaries. An unbiased sample of deep $X_{max}$ events in the energy interval
between $10^{18}$ to $10^{18.5}$ eV
was used to measure the cross section. 
In Fig. \ref{crossection} (left) the result of an unbinned maximum likelihood fit 
of an exponential function to the the tail of the $X_{max}$ distribution is shown. 
To properly account for shower fluctuations and detector effects, the exponential 
tail is compared to Monte Carlo predictions. Any disagreement between data
and predictions can then be attributed to a modified value of the
proton-air cross section\cite{prl}. 
The result, after averaging the values of the cross section for 
different hadronic interaction models, yields $\sigma_{p\,air}=505$ ($\pm22$ stat) ($+20-15$ syst) 
mb at a center-of-mass energy of $57\pm6$ TeV. 
The systematic uncertainty is dominated by mass composition(mainly the Helium fraction) 
and hadronic interaction models.
The result, shown in Fig. \ref{crossection} (right), favors a moderately slow rise of the cross-section towards higher energies, 
as observed at the LHC\cite{LHC}.
\section{Longitudinal profile and ground signal missmatch: muon content}
\begin{figure}[pb]
\centerline{
\includegraphics[width=6.cm]{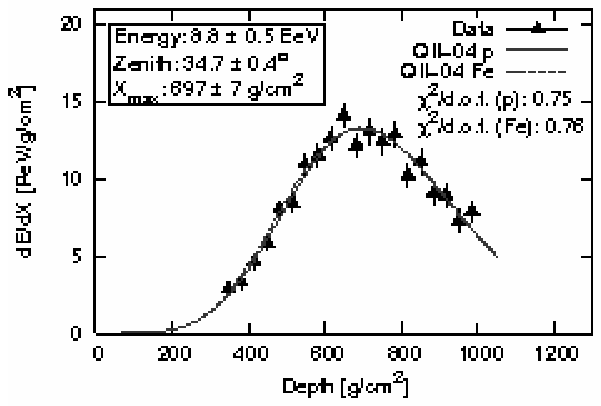}
\includegraphics[width=6.cm]{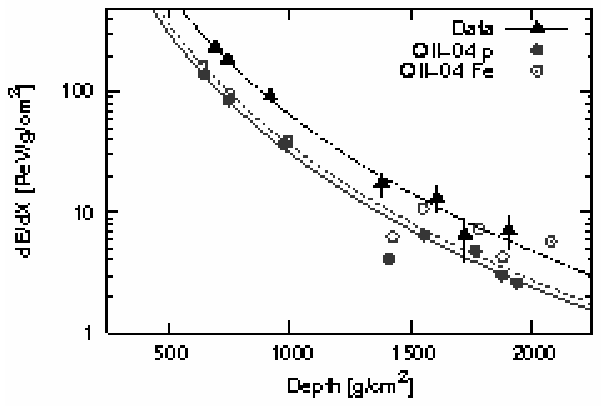}
           }
\vspace*{8pt}
\caption{(left)-Measured longitudinal profile of a typical air shower with two of its matching simulated air showers,
for a proton and an iron primary, simulated using {\sc QGSJETII-04}. 
(right)- The observed simulated ground signals for the same event.
\label{mismatch}}
\end{figure}
\begin{figure}[pb]
\centerline{
\includegraphics[width=6.5cm]{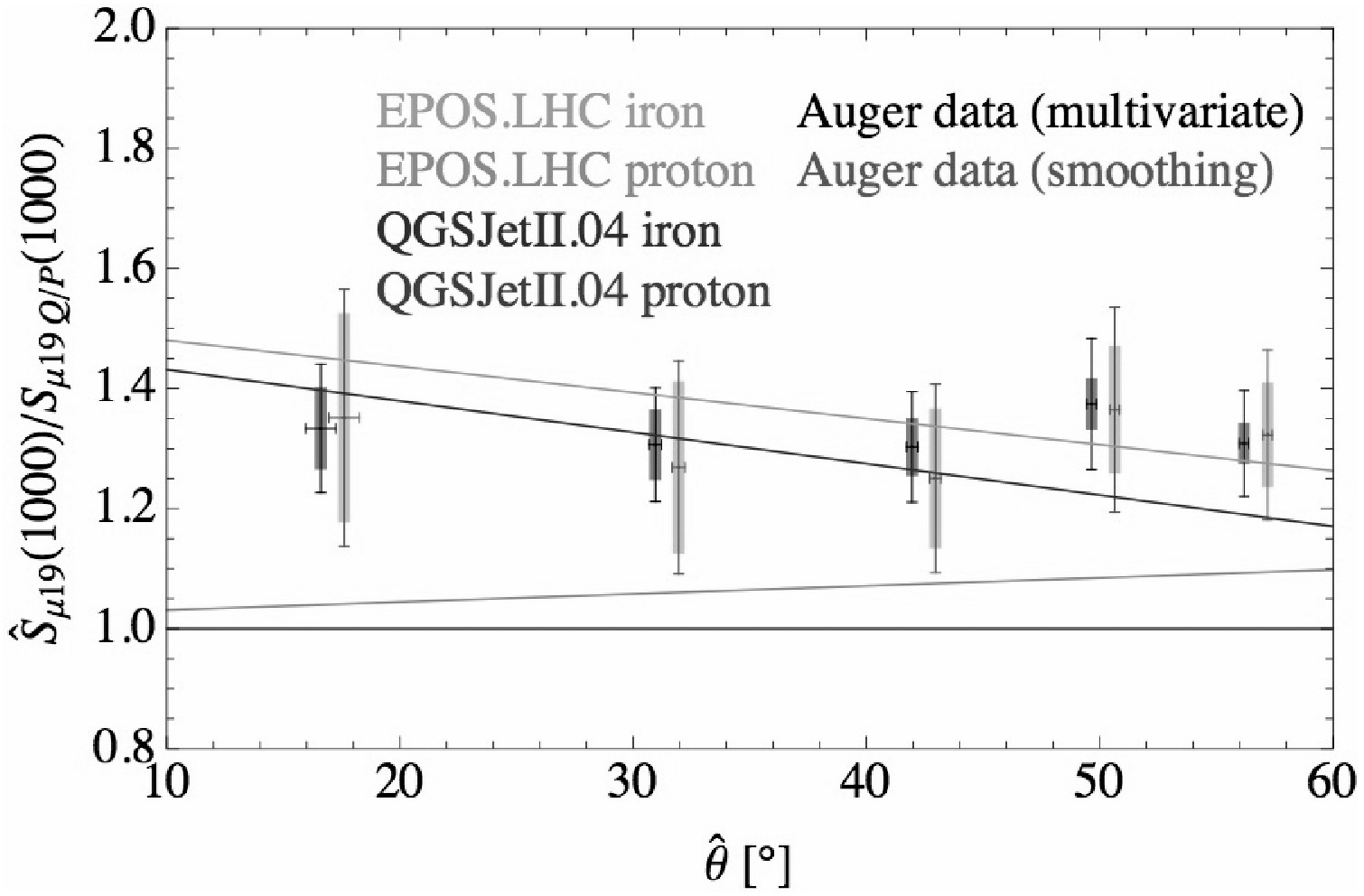}
\includegraphics[width=4.5cm]{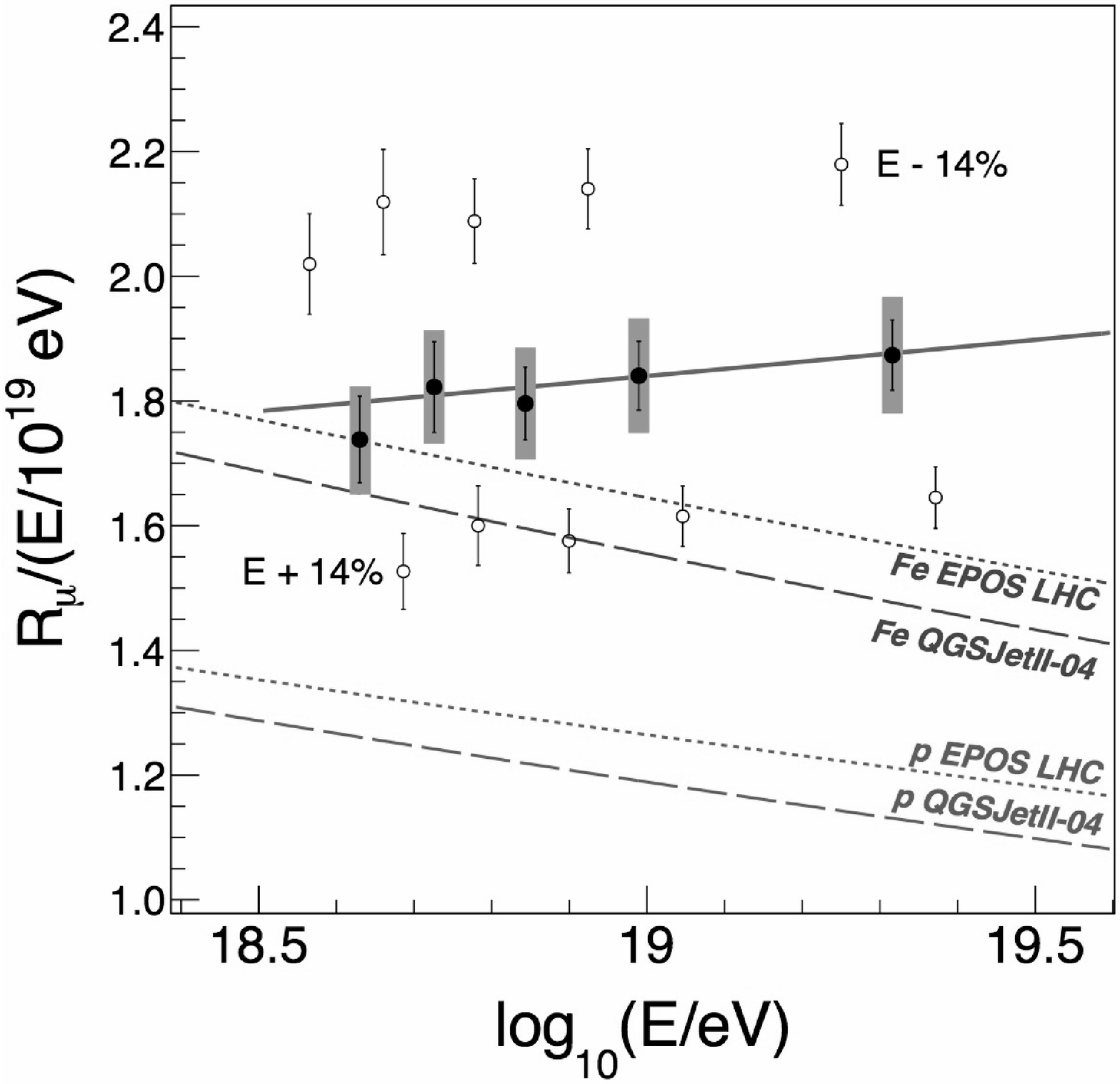}
           }  
\vspace*{8pt}
\caption{
(left) The measured muon signal at 1000 m from the shower axis vs. zenith angle, with respect to 
{\sc QGSJETII-04} proton at $10^{19}$ eV as baseline, obtained by different methods. 
(right) Number of muons from inclined showers compared to model predictions, as a function of energy.
\label{muon}
}
\end{figure}
Discrepancies were found in showers measured in hybrid mode,
when measurements of the longitudinal shower profile 
were compared to the lateral particle distributions at ground level\cite{icrc2013mismatch}. 
For each shower, Monte Carlo simulated events with similar energies 
were generated selecting those matching the measured longitudinal profile.
When the predicted lateral distributions of the signal are compared to the data recorded 
by the SD, the Monte Carlo predictions are found to be systematically below 
the observed signals, regardless of the hadronic model being used, and for all 
composition mixes that fit the $X_{max}$ distribution of the data sample (see Fig. \ref{mismatch}).

Independent methods of extracting muon content from very inclined showers\cite{icrc2013muonhorizontal} or 
relying on the different signal shape produced by muons in the water-Cherenkov
detectors\cite{icrc2013muonfadc}, together with this analysis of hybrid events,
indicate that there is a deficit of muons in the simulations when 
current hadronic interaction models are used, unless a pure iron primary composition 
is assumed. 
A pure iron composition would be in contradiction to the $X_{max}$ data 
when interpreted using the same models, leading to the conclusion that 
shower models do not correctly describe the muonic ground signals.
The muon content can be obtained from a SD signal time structure as muon signals stand 
above the smooth electromagnetic component. Results using 
the multivariate and smoothing methods are shown in Fig \ref{muon} (left).
Above a zenith angle of $62^{o}$, muons dominate the recorded signals at ground level,
as the electromagnetic component is absorbed in the atmosphere. 
A direct measurement of the muon content from very inclined
showers is shown in Fig. \ref{muon} (right).
\section{Conclusions}
The Pierre Auger Observatory can be used to test the properties of hadronic interactions 
by comparing simulations with data for independent observables. 

The proton-air inelastic cross section was extracted from 
measurements of the longitudinal development of air showers
at a centre of mass energy per nucleon of 57 TeV, 
well above energies of accelerator experiments. 

Deviations were found if the FD longitudinal profile and SD signals 
are compared in hybrid events. Independent methods using the SD indicate that 
this is due to a significant muon deficit in the predictions. 
A realistic treatment of the mass composition does not remove the muon discrepancy. 
Multiple methods reach the same conclusion: current hadronic interaction models 
do not accurately describe the muon signal. 

\end{document}